\def\puncspace{\ifmmode\,\else{\ifcat.\C{\if.\C\else\if,\C\else\if?\C\else%
\if:\C\else\if;\C\else\if-\C\else\if)\C\else\if/\C\else\if]\C\else\if'\C%
\else\space\fi\fi\fi\fi\fi\fi\fi\fi\fi\fi}%
\else\if\empty\C\else\if\space\C\else\space\fi\fi\fi}\fi}
\def\SP{\let\\=\empty\futurelet\C\puncspace}

\def\h1{$h^{-1}$\SP}

\def\etal{{\it et al.\/}\ }

\def\lsim{~\rlap{$<$}{\lower 1.0ex\hbox{$\sim$}}}
\def\gsim{~\rlap{$>$}{\lower 1.0ex\hbox{$\sim$}}}
\def\void#1{{}}

\def\xi2{$\chi^2$}

\documentclass{aaa}

\usepackage{graphicx,psfig,amsmath}

\newcommand{\less}{\raisebox{-1.1mm}{$\stackrel{<}{\sim}$}}
\newcommand{\more}{\raisebox{-1.1mm}{$\stackrel{>}{\sim}$}}
\newcommand{\msol}{{M$_{\odot}$}}
\newcommand{\csl}{{\sc class\_star\_lim}}
\newcommand{\cs}{{\sc class\_star}}
\newcommand{\msl}{{\sc mag\_star\_lim}}

\begin{document}

\title{ESO Imaging Survey\thanks{Full Figures 2, 4, 5, 6, 8 and 9 are only available 
in the on-line edition of A\&A, and Tables 3-8 are only available in electronic form
at the CDS via anonymous ftp to cdsarc.u-strasbg.fr (130.79.128.5)
or via http://cdsweb.u-strasbg.fr/cgi-bin/qcat?J/A+A/.} }
{\subtitle{The Stellar Catalogue in the Chandra Deep Field South}} 

\author{ 
M.A.T. Groenewegen\inst{1,2}, L. Girardi\inst{3,4}, E. Hatziminaoglou\inst{2}, 
C. Benoist\inst{5}, L.F. Olsen \inst{6}, L. da Costa\inst{2}, S. Arnouts\inst{2}, 
R. Madejsky\inst{2,7}, R.P. Mignani\inst{2}, C. Rit\'e\inst{2,8}, G. Sikkema\inst{2}, 
R. Slijkhuis\inst{2} \and B. Vandame\inst{2}
}

\institute{
Instituut voor Sterrenkunde, PACS-ICC, Celestijnenlaan 200B, B--3001 Heverlee, Belgium
\and
European Southern Observatory, Karl-Schwarzschild-Stra{\ss}e 2, D--85748
Garching b. M\"unchen, Germany
\and 
Osservatorio Astronomico di Trieste, Via Tiepolo 11, I--34131 Trieste, Italy
\and
Dipartimento di Astronomia, Universit\`a di Padova, Vicolo dell'Osservatorio 5, I--35122 Padova, Italy
\and 
Observatoire de la C\^ote d'Azur, BP 229, F--06304 Nice, cedex 4, France
\and 
Astronomical Observatory, Juliane Maries Vej 30, DK--2100 Copenhagen, Denmark
\and 
Universidade Estadual de Feira de Santana, Campus Universit\'ario, Feira de Santana, BA, Brazil
\and 
Observat\'orio Nacional, Rua Gen. Jos\'e Cristino 77,  Rio de Janerio, R.J., Brasil
}

\date{Received, accepted}

\offprints{
groen@ster.kuleuven.ac.be 
}

\authorrunning{Groenewegen et al.}
\titlerunning{The Stellar Catalogue in the Chandra Deep Field South.}

\abstract{ 
Stellar catalogues in five passbands ($UBVRI$) over an area of
approximately 0.3 deg$^2$, comprising about 1200 objects, and in seven
passbands ($UBVRIJK$) over approximately 0.1 deg$^2$, comprising about
400 objects, in the direction of the Chandra Deep Field South are presented.
The 90\% completeness level of the number counts is reached at
approximately $U$ = 23.8, $B$ = 24.0, $V$ = 23.5, $R$ = 23.0, $I$ =
21.0, $J$ = 20.5, $K$ = 19.0.
These multi-band catalogues have been produced from publicly available,
single passband catalogues.  A scheme is presented to select point
sources from these catalogues, by combining the SExtractor parameter
\cs\ from all available passbands.
Probable QSOs and unresolved galaxies are identified by using the
previously developed $\chi^2$-technique (Hatziminaoglou \etal 2002), that
fits the overall spectral energy distributions to template spectra and
determines the best fitting template. Approximately 15\% of true
galaxies are misclassified  as stars by the $\chi^2$ method.
The number of unresolved galaxies and QSOs identified by the
$\chi^2$-technique, allows us to estimate that the remaining level
of contamination by such objects is at the level of 2.4\% of the
number of stars. The fraction of missing stars being incorrectly
removed as QSOs or unresolved galaxies is estimated to be similar.
The observed number counts, colour-magnitude diagrams, colour-colour
diagrams and colour distributions are presented and, to judge
the quality of the data, compared to simulations based on the
predictions of a Galactic Model convolved with the estimated
completeness functions and the error model used to describe the
photometric errors of the data. By identifying outliers in specific
colour-colour diagrams between data and model the level of
contamination by QSOs is alternatively estimated to be $\less$ 6.3\% in
the seven passband and \less 2.3\% in the five passband catalogue. This,
however, depends on the exact definition of an outlier and the
accuracy of the representation of the colours by the simulations.
The comparison of the colour-magnitude diagrams, colour-colour
diagrams and colour distributions show, in general, a good agreement
between observations and models at the level of less than 0.1 mag;
the largest discrepancies being a colour shift in $V-R$ and $R-I$ of
order of 0.2 mag possibly due to uncertainties in the bolometric corrections.
Although no attempt is made to fit the model to the data, a comparison
shows that the lognormal law for the initial mass function proposed by
Chabrier (2001) describes the data better than the power law form in
that paper.
The resulting stellar catalogues  and the objects identified as likely
QSOs and unresolved galaxies with coordinates, observed magnitudes
with errors and assigned spectral types by the $\chi^2$-technique are
presented and are publicly available.
}


\maketitle


\section{Introduction}
\label{intro}

The primary goal of the ESO Imaging Survey project is to conduct
public imaging surveys to provide data sets from which statistical
samples of different types as well as of rare population objects can
be drawn for follow-up observations with the VLT. A summary of the
surveys conducted so far can be found in da Costa \etal (2001). 
Recently, deep optical/infrared observations of the Chandra
Deep Field South (CDF-S) have been completed as part of the ongoing
Deep Public Survey (DPS), and the data publicly released (Vandame \etal
2001; Arnouts \etal 2001). In these papers, fully calibrated images and
single passband catalogues were presented for $UBVRI$ observations
covering an area of 0.3 deg$^2$ and for $UBVRIJK$ over a region of 0.1
deg$^2$ at the centre of the former area.  Even though the angular
coverage is still relatively small, the particular combination of
depth and solid angle of the complete survey and the uniformity of the
multi-band data which will result make it worthwhile to explore the
type of objects likely to be sampled by the survey, to devise analysis
techniques to identify objects of potential interest and to make a
first evaluation of the samples that are likely to emerge when the
full survey covering 3 deg$^2$, is completed.

The present work is in many ways an extension of that presented in
Hatziminaoglou \etal (2002; hereafter H2002) which used a
$\chi^2$-technique to classify point sources according to spectral
type using their multi-band photometric properties, trying to identify
special objects such as QSO, White Dwarf (WD) and low mass star candidates.

Here, the question of how to choose an optimum stellar catalogue for
statistical studies, either using a single passband or using
multi-colour data, is discussed in more detail.  Procedures are
developed to set the contamination by extended sources below a
specified value at the single passband level. To reach fainter
magnitude limits the leverage of using multi-colour data is
investigated. Finally, to assess to some extent the quality of the
final multi-colour stellar catalogue produced, number counts, colour
distribution, and colour-colour and colour-magnitude diagrams are
compared to those predicted by a Galactic model calibrated using
independent data.

In Section~\ref{data} the data are briefly reviewed as well as the
method employed in the construction of the point source catalogue used
in the present paper. Section~\ref{SPS} discusses how point sources
are selected in the single passband, and colour catalogues, and how
potential unresolved galaxies and QSOs are found using a previously
developed $\chi^2$-technique. In Section~\ref{Procedure} the actual
procedure for the creation of the single passband to the final stellar
colour catalogue is presented. Section~\ref{GM} and Section~\ref{MS}
introduce the Galactic Model and the simulation of appropriate data
sets.  Section~\ref{Results} presents the results.  Finally, in
Section~\ref{summary} a brief summary is presented.

\section{The data}
\label{data}

The single passband catalogues for the CDF-S are taken from Vandame
\etal (2001) and Arnouts \etal (2001), cut at a $S/N=2$, slightly lower
than the catalogues publicly released which were cut at a $S/N=3$. 

This is done as a magnitude, even with a larger error bar, is
preferred over an upper limit which would otherwise be assigned to the
magnitude in a filter where an object is not detected.

The $U$-band catalogue used here (and in H2002) was extracted from an
image produced by stacking all the available $U/38$ and $U350/60$ images,
to reach a fainter limiting magnitude. A more detailed
discussion of this catalogue and of its photometric calibration will
be present in Arnouts \etal (in preparation).  The main conclusion is
that a shift of +0.2 mag has to be applied to the 
$U$-magnitudes\footnote{All magnitudes in the present paper are in the
natural system of the WFI filters at the ESO/2.2m telescope for
$UBVRI$, and the SOFI filters at the ESO/NTT for $JK$, with Vega based
zero points.} given by SExtractor running on the stacked $U$ image using the
photometric zero point of the $U/38$ image, to get agreement
between observed and predicted $U-B$ colours for normal stars.

To build a colour catalogue, the single passband catalogues
are associated on position using, in the present paper, a fixed search
radius of 0.8\arcsec. With this search radius the number of multiple
associated sources is kept low ($\less$0.8\% in all passbands, and
typically 0.2\%), while resulting in many positive matches. A detailed
discussion on the optimum search radius, and more complicated schemes
using variable search radii will be presented elsewhere (Benoist \etal 2002).

After the association, only objects that are in the area common to all
catalogues and outside all the masks placed around saturated objects
and bright stars are kept. This ensures that all objects could, at
least in principle, have been detected in all passbands.  The
effective area is 0.263 deg$^2$ for the five passband catalogue and
0.0927 deg$^2$ for the seven passband catalogue.
If an object is not associated in a passband, an 3$\sigma$ upper limit is
assigned for the magnitude in that passband.

\section{Selecting Point Sources}
\label{SPS}

\subsection{Selecting Point Sources at the single passband level}

\begin{figure}[ht]
\centerline{
\psfig{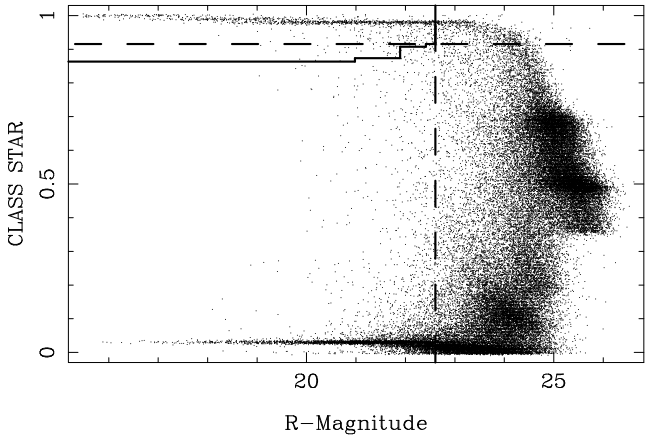}}
\caption{
SExtractor parameter \cs\ versus magnitude for the 
$R$-band, with the finally adopted morphological classification limits
indicated by the solid line.  Objects in the upper left box are
considered point sources in a single passband.  The dashed line
indicates the much simpler criterion used previously which only took
into account the \cs\ value at the faintest \msl.
}
\label{fig:CS}
\end{figure}

The SExtractor software package used at the moment to detect and
extract sources uses a neural network to separate point from extended
sources. For each object it returns a parameter, \cs, with a value
between 1 (a perfect point like object) and 0.  An example of the
distribution of \cs\ as a function of magnitude is shown in
Fig.~\ref{fig:CS} for the $R$-band.

The neural network file used here is the standard one that is part of
the SExtractor package and that has been trained for seeing FWHM values
from 0.025 to 5.5\arcsec\ and for images that have 1.5 $<$ FWHM $<$ 5
pixels, and is optimised for Moffat profiles with 2$\le \beta \le$ 4.
These conditions are fulfilled for the EIS optical and IR observations
of the CDF-S.

At the bright end of the distribution shown in Fig.~\ref{fig:CS} two
sequences can easily be distinguished with \cs\ values below 0.1 and
above 0.95, respectively. However, for fainter magnitudes the
sequences spread out and merge. In previous EIS papers (e.g. Prandoni
\etal 1999) the notion of a morphological classification limit \msl\
was introduced, where objects brighter than this limit and with a \cs\
value larger than a limit \csl\ were considered point sources. In
previous papers, these limits were chosen basically by eye from
figures like Fig.~\ref{fig:CS}.

Clearly, there is a need to determine \csl\ and \msl\ in a more
systematic way, by statistically describing the relative contribution
of the point- and extended-source population.  The goal is to produce
a magnitude-dependent classification as illustrated in Fig.~\ref{fig:CS}.

A  technical description of the actual implementation of the
concept used in the present paper is given in the Appendix, and
Fig.~\ref{fig:CS} shows for the $R$-band the distribution over \cs\
and the adopted values for \csl\ and \msl\ (solid line) in the present
paper and for comparison the old way of using a single value (dashed line).
It should be noted that this procedure allows the estimate of the
smallest \cs\ where stars can be distinguished over the large
background of galaxies. It does not imply, especially at the faint
end, that stars with \cs\ values lower than \csl\ do not exist. 
Furthermore, it is recalled that the actual distribution of
\cs\ versus magnitude depends on the seeing but also on the 
Galactic coordinates of the field under consideration. 

It is important to emphasise that more work should be done to explore
alternative ways of morphologically classifying images. One
alternative under consideration is to use multi-resolution methods to
provide an independent pixel-based classification scheme (Vandame,
private communication).


Apart from a statistical analysis of the distribution of \cs\ versus
magnitude a second criterion of selecting point sources is
investigated, namely the fact that they are expected to be ``round''.
This is done by considering the SExtractor parameters $A$ and $B$ 
which represent the profile fitting rms along the major and
minor axis 
(see Arnouts \etal 2001 for more discussion). 
Visual inspection of some objects with large values of
$\mid(A-B)/(A+B)\mid$ showed that these outliers are not stellar, but
bad pixels, or objects close to the edge of a frame or a mask.  The
standard deviation of the distribution is computed and objects that
deviate by more than a specified level (in the present paper
6$\sigma$) are not considered point sources.  Typically, 0.1\% of the
objects are removed in this way. A 3$\sigma$ clipping, however was
verified to be too stringent, as then many true stars would be
removed. Most of them would be recovered by the colour Point/Extended
(hereafter, P/E) source classification scheme discussed below, but
they would, incorrectly, be excluded from the single passband number counts.

\subsection{Selecting Point Sources in a colour catalogue}

The (P/E) source classification in a colour catalogue is a non trivial
matter, as the \msl\ and \csl\ of all the passbands involved must be
combined in some way. Here the following scheme is adopted.  As the
(P/E) source classification is most reliable for bright objects (see
Fig.~\ref{fig:CS}), the classification in the colour catalogue will
be determined by the passband in which the object is brightest
relative to the (faintest) \msl\ in that passband.  If the object is
classified as a point source in this passband, it is considered a
point source in the colour catalogue. The fact that the seeing in the
images is different is implicitly taken into account by the fact that
the \msl\ in a case of worse seeing is brighter than in the case of good seeing.

A completely different way of creating a colour catalogue is by
digitally stacking the images in the various filters, and then running
the source detection algorithm on it. The photometry is then obtained
by going back to the individual filters and measure the flux around
the now known positions. This method was applied to EIS data by da
Costa (1998) using the algorithm of Szalay et al. (1999). From that
work, SExtractor's standard neural network for (P/E) source
classification is known not work properly on such a stacked image
because of e.g. seeing variations among the different
filters. Therefore this procedure is not further pursued here.

\subsection{Eliminating  non-stellar objects}

The procedures outlined above can only separate point-like from
extended objects.  However, to produce a truly stellar catalogue one
has to consider the possibility of contamination by QSOs and unresolved
galaxies.  This is not possible in a single passband catalogue. However, in a 
multi-colour catalogue the photometric data provides information on the
shape of the Spectral Energy Distribution (SED) which can, in
principle, be used to address this issue.

In fact, this was the subject of the paper of H2002 where a colour
point source catalogue was used to classify sources according to their
spectral type. This was done by $\chi^2$-fitting template spectra to
the observed magnitudes in different bands. The spectral library in
use consists of series of: model QSO, WD and brown dwarf spectra;
three empirical cool WD spectra; a set of stellar templates for O-M
stars; and a set of model galaxy spectra. In H2002 emphasis was given
in identifying QSOs, WD and low mass star candidates for follow-up
spectroscopy.
The $\chi^2$-technique provides an independent view of the nature of a
source, complementing the morphological classification. However, by
using broadband photometric data to assign spectral types,
misclassifications are possible (e.g. due to the fact that an
object is an unresolved binary leading to composite colours) and this
is further discussed in Sect.~\ref{effectiveness}.

\subsection{Choosing parameters}

The morphological classification limits adopted in the present paper
are listed in Tab.~\ref{tab:cllim} and are chosen after extensive
testing.  In terms of the technical description in the Appendix the
parameter $t_2$ is set to 30 and the parameter $t_1$ to 80.

The main guideline to chose these particular values for $t_1$ and
$t_2$ to set the \msl\ and \csl\ values is the estimate of the
remaining contamination by QSOs and unresolved galaxies, and related,
the incorrect removal of true stars by misclassification of the
$\chi^2$-technique, which one wants to keep small (of order a few
percent at most).  These numbers are estimated independently in
Sects.~\ref{outliers} and \ref{effectiveness}.

\begin{table*}[ht]
\caption{Morphological classification limits \msl\ ($m_{\rm x}$) and \csl\ ($c_{\rm x}$) adopted 
in the single passband catalogues (see Fig.~\ref{fig:cs_c} for a visual representation).}
\label{tab:cllim}
\begin{tabular}{ccccccccc} \hline

Passband & $c_1$ & $m_1$ & $c_2$ & $m_2$ & $c_3$ & $m_3$ & $c_4$ & $m_4$  \\
\hline
%
%
 $U$ &       &       & 0.851 & 21.57 & 0.873 & 22.27 & 0.952 & 22.45 \\
 $B$ &       &       & 0.848 & 23.04 & 0.933 & 23.75 & 0.961 & 23.84 \\
 $V$ & 0.590 & 21.15 & 0.828 & 22.15 & 0.879 & 22.68 & 0.935 & 22.69 \\
 $R$ & 0.863 & 20.98 & 0.873 & 21.89 & 0.907 & 22.41 & 0.915 & 22.61 \\
 $I$ & 0.603 & 19.30 & 0.663 & 20.25 & 0.744 & 20.77 & 0.885 & 21.09 \\
 $J$ &       &       & 0.644 & 19.32 & 0.929 & 20.18 & 0.923 & 20.36 \\
 $K$ &       &       &       &       & 0.750 & 17.71 & 0.943 & 18.46 \\
\hline
\end{tabular}
\end{table*}

\begin{table}[ht]
\caption{Number of objects in the five and seven passband  point source catalogues}
\label{tab:infl}
\begin{tabular}{lcc} \hline
                                    & $UBVRI$   \\ \hline
Number of point sources             & 1371                 \\
Passband that decided (UBVRI)       & 7/308/0/650/406      \\
Not classified  by $\chi^2$ method  &   18                 \\
    Classified  by $\chi^2$ method  & 1353                 \\
\hspace{5mm} as rank 1              &  947                 \\
\hspace{5mm} as rank 2              &  127                 \\
\hspace{5mm} as rank 3              &  279                 \\
Number of unresolved galaxies       &   37                 \\
Number of QSOs                      &  136                 \\
Number of stars                     & 1198                 \\
\hline
                                    & $UBVRIJK$   \\ \hline
Number of point sources             & 486                   \\
Passband that decided (UBVRIJK)     & 2/110/0/225/37/110/2  \\
Not classified  by $\chi^2$ method  &   5                   \\
    Classified  by $\chi^2$ method  & 481                   \\
\hspace{5mm} as rank 1              & 160                   \\
\hspace{5mm} as rank 2              & 117                   \\
\hspace{5mm} as rank 3              & 204                   \\
Number of unresolved galaxies       &  16                   \\
Number of QSOs                      &  61                   \\
Number of stars                     & 409                   \\

\hline
\end{tabular}
\end{table}

\section{The procedure}
\label{Procedure}

For the chosen combination of \csl\ and \msl, the single passband
catalogues are produced and then associated to form the colour
catalogue with the (P/E) source classification applied based on all
the available colours. This is done separately for the $UBVRI$ and
$UBVRIJK$ catalogues. The resulting number of point sources is given
as the first entry in Table~\ref{tab:infl}. The second entry gives
the split over the passband that actually decided that a source is
considered a point source. In the five passband catalogue this is
mainly $R$, $I$ and $B$; in seven passbands mainly $R$, $J$ and $B$.

As a second step the $\chi^2$ template fitting of H2002 is applied to
the objects classified as point sources. This is done twice: once with
and once without the galaxy template spectra included.  The outcome of
the template fitting is the spectral type of the best fitting template
spectrum (with associated photometric redshift in the case of QSOs and
galaxies), the (reduced) $\chi^2$ and the ``rank'' (which is based on
the $\chi^2$ value, rank 1 being the robust candidates, rank 3 poor
candidates).  Objects that are ranked 3 when no galaxy templates are
considered (typically having reduced $\chi^2$ larger than 4-7), and
become ranked 1 galaxies when galaxy templates are considered, that
are fainter than 16th magnitude (in the passband where they are
brightest) and with photometric redshifts larger than 0.25 are
considered to be (unresolved) galaxies and are removed from the list
of point sources.  The split over the assigned ranks, the number of
unresolved galaxies and QSO, and the number of objects considered true
stars are given in Table~\ref{tab:infl}.

The $\chi^2$-technique requires an object to be detected in at least
three passbands. Less than 1.5\% of objects does not fulfil this
criterion (see the third entry in Tab.~\ref{tab:infl}). They are kept
as stars, as they are originally classified as point sources and there
is no evidence to the contrary. Exclusion of this small number of
objects does not affect any of the conclusions of this paper.

From an observational point of view, the validity of the adopted
approach to remove the most likely galaxy and QSO interlopers can be
verified only by spectroscopic follow-up of a representative sample of
objects. From a theoretical point of view it is planned in the future
to simulate the physical size, redshift and colour space occupied by
galaxies and QSOs as well, so it can be verified if a unresolved
source within a given colour bin is more likely to be a star, galaxy
or a QSO.

The final number of objects considered to be stars is about 1200
over the 0.263 deg$^2$ for the five passband catalogue and about 400
over the 0.0927 deg$^2$ for the seven passband catalogue. For
comparison, the number of extended objects over these areas is about
74000 and 28000, respectively.

\section{The Galactic  model}
\label{GM}

The observed counts and colours will be compared to  a newly
developed Galactic model (Girardi \etal in preparation).  In brief,
this model is based on a population synthesis approach, where stars
are randomly generated from an extended library of evolutionary tracks
according to user defined star formation rate (SFR), age-metallicity
relation (AMR), and initial mass function (IMF), and then distributed
in space according to the probabilities given by the specified
geometry. The evolutionary tracks include brown dwarfs down to 0.01
\msol, and evolved phases with the complete white dwarf stage for
stars up to 7 \msol. For the present paper the following ingredients
have been adopted:

\begin{itemize}

\item The IMF is taken to be the lognormal form proposed by Chabrier (2001). 
The effect of adopting the power law form proposed by him will be
detailed in Sect.~\ref{cIMF}

\item The SFR is assumed to be constant over the last 11 Gyr 
for the disk, and constant between 12 and 13 Gyr for the halo.

\item The AMR for the disk is taken from Rocha-Pinto \etal (2000). 
[Fe/H] values are converted into the metal content $Z$ by means of a
relation that allows for $\alpha$-enhancement at decreasing [Fe/H], as
suggested by Fuhrmann (1998) data. At any age, [Fe/H] is assumed to
have a 1$\sigma$ dispersion of 0.2 dex.

The metallicity of the halo stars is assumed be $Z$ = 0.0095, with a
dispersion of 1.0 dex. This is based on an observed [Fe/H] value of
$-1.6 \pm 1.0$ (Henry \& Worthey 1999), allowing for an
$\alpha$-enhancement of 0.3 dex.

\item The disc component is described by a double-exponential in 
scale height and Galactocentric distance.  The model does not have
separate components for thin or thick disk.  Instead the scale height
for disk stars is a function of age, and is parameterised as:
\begin{equation}
  H = z_0 \;\; (1 + t / t_0)^{\alpha}
\end{equation}
with $t$ the age of the star.  Rana \& Basu (1992) for example find
$z_0$ = 95 pc, $t_0$ = 0.5 Gyr and $\alpha$ = (2/3).  However this
does not fit very well the derived scale height of `thick', `old',
`intermediate' and `young' disk components as derived by Ng \etal
(1997). Their results are described by $z_0$ = 95 pc, $t_0$ = 4.4 Gyr
and $\alpha$ = 1.66, which are adopted here.

\end{itemize}

For the present paper the (relative) local densities of halo and disk
stars, the oblateness of the halo and the Galactocentric disc scale
length are derived from the ``Deep Multicolor Survey'' (DMS; Osmer
\etal 1998 and references therein) covering six fields and EIS data
(Prandoni \etal 1999) for the South Galactic Pole. A more detailed
discussion of the calibration of the Galactic model will be given in
Girardi \etal (in preparation). Note that the Galactic model
parameters derived below differ slightly from those used by H2002 to
compare some of their results to.

The halo oblateness (and local halo number density) is derived by
fitting the number of halo stars, defined by (in the Johnson-Cousins
system) $0 < B-V < 0.7$ in the range $20 < B < 22$ and $0 < V-I < 0.8$
in the range $18 < I < 20$, in these seven fields and is found to be
$q = 0.65 \pm 0.05$. This value is smaller than the value of 0.8 $\pm$
0.05 quoted by Reid \& Majewski (1993), but recently Robin \etal
(2000) could not exclude a spheroid with a flattening as small as $q =
0.6$ and Chen \etal (2001) derived $q = 0.55 \pm 0.06$.

The disc radial scale length (and local disk number density) is
derived by fitting the number of disk stars, defined by $1.3 < B-V <
2.0$ in the range $20 < B < 22$ and $1.8 < V-I < 4.0$ in the range $18
< I < 20$, and is found to be $H_{\rm R} = 2800 \pm 250$ pc. This is
in agreement with the lower limit of 2.5 kpc (Bahcall \& Soneira 1984)
and the recent work of Zheng \etal (2001) on M-dwarfs who derive
$H_{\rm R} = 2750 \pm 160$ pc and Ojha (2001) who derive $H_{\rm R} =
2800 \pm 300$ pc for the thin disc.

The model with these parameters is applied to the CDF-S field. The
Galactic bulge has no contribution in this direction. The model is
calculated for an area of 3 deg$^2$, which is about 10 times larger
than the actual area covered by the $UBVRI$ data, and about 30 times
larger than the area covered by the $UBVRIJK$ observations.  This
ensures that the parameter space is properly sampled and that the
errors on the model data are at least 3, respectively 5, times
smaller than the error bars on the observed data.

\section{Simulations of the observed data}
\label{MS}

The galactic model produces ``perfect'' stellar objects down to a
specified magnitude (in fact the model is computed down to $R$ = 30.5). 
To build a mock colour catalogue simulating the real data, the
magnitudes from the galactic model are taken as input and the
following steps are considered:

\begin{itemize}

\item The error in the magnitude (as computed by SExtractor) as a 
function of magnitude for each band (based on the properties of the
actual image and real data). Additionally, one can scale these errors
by a factor and add in quadrature an ``external error'', simulating
for example errors in the photometric zero point. The ``observed''
magnitude is then drawn from a Gaussian with the estimated error as
width and centred around the ``perfect'' magnitude. The scaling
factor and ``external error'' used in the present paper are given below.

\item The saturation at bright magnitudes. The ADU level in the central 
pixel of an object is simulated, and compared to the actual value used
by SExtractor to flag a saturated object.

\item The completeness at the faint end. Artificial star tests using 
the EIS images themselves have not yet been performed to estimate the
completeness functions. As a temporary solution the completeness
functions have been derived for the $R$-band and $K$-band by comparing
the EIS data to the deeper observations of respectively Wolf \etal
(2001) and Saracco \etal (2001) over the parts where their
observations overlap with the EIS observations.  For the other bands
the shape of the function that describes the completeness is kept the
same (for $UBVI$ taking $R$ as a reference, and for $J$ taking $K$ as
the reference), and scaling the characteristic magnitude where the
incompleteness sets in using the 5$\sigma$ limiting magnitude of the
bands as a reference. For the stellar catalogue considered here the
exact details of the completeness functions are not of too much
importance as the estimated magnitudes where the EIS data is estimated
to be 50\% complete ($U$ = 25.1, $B$ = 26.6, $V$ = 25.5, $R$ = 25.4,
$I$ = 24.3, $J$ = 22.9, $K$ = 21.2) are at the very faint end of the
stellar catalogue.  In other words, the depth of the stellar catalogue
is set by the ability to separate point from extended sources and the
remaining contamination by QSOs and unresolved galaxies, not the depth
or incompleteness of the observations as such.
\\
\noindent These same observations are used to compare the differences in the
magnitudes of the objects in common between EIS and the external
data. Slices in magnitudes are made, and the average error for the EIS
observations, the average error for the comparison catalogue (hence Wolf
\etal for $R$, and Saracco \etal for $K$), and the rms of the
difference of the stars in common determined. The errors in the
external catalogues are always smaller than in the EIS data, as
expected since these observations are deeper than ours. However, both
underestimate the error as compared to the rms in the difference.
Based on these comparisons an external error of 0.01 mag in $UBVRI$
and 0.03 mag in $JK$, and a scaling factor of two in all passbands are
adopted in the simulation of the errors. With this description the
width of the stellar sequence in the colour-colour diagrams is
described extremely well for all magnitudes and passbands (see
Fig.~\ref{fig:CC_dia}).

\end{itemize}

\noindent The simulation also takes into account the S/N limits
imposed on the single passband and colour catalogues.  Furthermore,
objects that are saturated in one or more bands are removed from the
mock catalogue, as masks are placed around them in the real data and
only objects outside all masks are being considered in the final
colour catalogue.

Finally, two types of mock stellar catalogues are produced. One
considering all simulated point sources that are potentially
``observable'', and the second catalogue that also takes into account
the way that point sources are being selected from the real colour
catalogue by checking if the object is brighter than the morphological
classification limit used in the production of the observed data set
in at least one passband (see Tab.~\ref{tab:cllim}).  This is a
conservative estimate, as for the real data the value of the
SExtractor parameter \cs\ also plays a role. However this feature in
fact allows us to check if the
\cs\ values chosen are not too low (resulting in too many galaxies and
leading to high stellar number counts at the faint end) or too
conservative (leading to number counts that are below the predictions).  
Details on the simulation of the photometric errors, the completeness
model and the simulated datasets will be given elsewhere (Groenewegen
\etal 2002, in preparation).

\section{Results and analysis}
\label{Results}

In this section the observed number counts, colour-colour and
colour-magnitude diagrams, and colour-distribution histograms are
presented and compared to the simulations. The effectiveness of using
the $\chi^2$-technique is discussed. Finally, catalogues of stellar
sources, QSOs, and unresolved galaxies are presented. Due to space
constraints only representative figures will be shown and the full set
of figures is only available through the on-line edition of A\&A.
Note that most of the figures need to be viewed or printed in colour.

\subsection{Number counts}
\label{numbercounts_c}

\subsubsection{General description}

\begin{figure}[ht]
\centerline{
\psfig{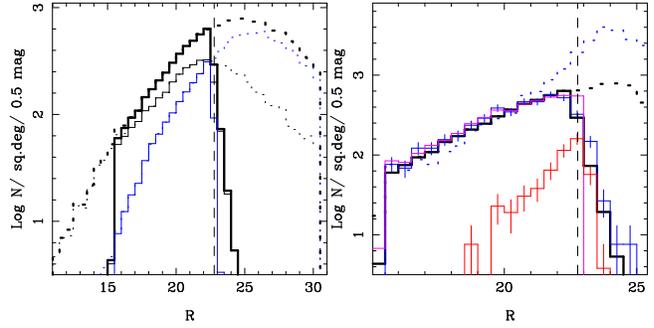}}
\caption{
Number counts in $UBVRI$ from the five passband catalogue and in $JK$
from the seven passband catalogue. This sample is for the $R$-band,
and the full figure is only available in the on-line edition.  In the lower right
panel the $R$-band counts based on the seven passband catalogue are
presented. They are in good agreement with those based on the five
passband catalogue, as is the case for $UBVI$.
Left panel: shown are the disk component (dashed and solid thin black
histograms) and halo component (dashed and solid blue histograms), and
their sum (dashed and solid thick black histograms) predicted by the
galactic model.  Distinguished are the predicted model counts taking
into account the error model, saturation, completeness and the \msl\
in the individual passbands (the blue and black full histograms), as
well as the ``perfect'' model (the blue and black dashed histograms).
The morphological classification limits for the bands are indicated by
the vertical dashed line.
Right panel: comparison of the data (blue coloured histogram with
errors bars) with the model. Note the different magnitude scale
w.r.t. the left-hand panel. Shown are the predicted counts taking into
account the error model, saturation and completeness (thick dotted
black line), as well as the counts that additionally take into account
the \msl\ limits in the individual passbands (thick solid black line;
the same curve as in the left-hand panel).  The pink histogram
represents the observed number counts using the point source
classification for that band only. The number counts of objects
assigned ``stellar'' template spectra that are ranked 1 and $\le3$ in
the full five and full seven passband catalogue respectively, based on
the $\chi^2$-technique only, are indicated by the dashed blue
histogram. The QSO numbers counts (among the objects initially
classified as point sources) are indicated by the red histogram.  The
morphological classification limits for the bands are indicated by the
vertical dashed line.
}
\label{fig:NC}
\end{figure}

Figure~\ref{fig:NC} shows the number counts for every passband from
the five passband catalogue for $UBVRI$ and from the seven passband catalogue
for $JK$. In the lower right panel the $R$-band number counts are shown
based on the seven passband catalogue. The agreement with that from the five
passband catalogue is excellent, and the same applies when comparing
the number counts in the other optical bands as derived from the five and
seven passband catalogues. 

In the left-hand panels the predictions by the galactic model are
shown, namely the contributions by the disk (dashed and solid thin
black histograms), the halo (dashed and solid blue histogram) and
their sum (dashed and solid thick black histogram).  Shown are the
predictions from the ``perfect'' model (the blue and black dashed
histograms), and the predictions taking into account the error model,
saturation, completeness and the \msl\ limits that are used to create
the colour catalogue (the blue and black full histograms). The
left-hand panels illustrate the relative contributions of disk and
halo at various magnitudes. In this particular field the contribution
of the disk exceeds that of the halo for almost all observed
magnitudes. The model predicts, however, that obtaining deeper images
and assuming one would be able to do the (P/E) classification at the
level of $J, K \approx 23-24$ the number counts should be dominated by
the halo. In the optical the halo dominates the disk only at much
fainter magnitudes ($I \approx 25$, $V \approx 28$). Deeper infrared
data could therefore better constrain the IMF and other parameters
describing the halo than the present data.

In the right-hand panels, the model predictions (again, the full thick
black histogram) are compared to various datasets. The blue histogram
with the error bars (based on $\sqrt{N}$ statistics) represents the
observed stellar number counts from the colour catalogue based on the
(P/E) classification and subsequent application of the $\chi^2$-technique
described in Sect.~\ref{Procedure}.  The pink histogram represents the
number counts using the (P/E) classification for that band {\it
only}. As, by construction, there are no stars fainter than \msl\ in
the single passband catalogue, but this value usually does not
coincide with the borders of the binning, the counts in the last bin
may drop artificially. Therefore, the counts in the last bin of the
pink histogram are corrected for this effect by multiplying the counts
in the last bin by the bin width divided by the actual width over
which there is data.

\subsubsection{Comparing counts from single passband and colour catalogues}

Up to \msl, indicated by the vertical dashed line, the counts from the
single passband and the colour catalogue are in very good agreement. 
In fact, naively, one might even have expected them to be identical.
They are not for several reasons.

Firstly, in particular in $U$, at the bright end, the counts from the
single passband are above the counts from the colour catalogue. This
is due to the fact that only objects which are outside {\it all} masks
created at the single passband level are considered in the colour
catalogue. Since the $U$-band is shallow, the single passband
catalogue contains objects that are not saturated in $U$, but are
saturated (hence will have a mask placed around them) in one of the
other bands, and will not make it into the colour catalogue.

Secondly, the ``good'' area over which the sources are considered is
slightly larger for the single passband catalogues compared to the
colour catalogue, as the different single passband images do not
exactly have the same field centre and the colour catalogue is created
over the area that is in common to all passbands. At bright and
intermediate magnitudes the counts match to within the Poisson errors,
as expected (except for the $U$-band for reasons just discussed).

Thirdly, when approaching the morphological classification limit the
number counts based on the colour catalogue tend to fall below the
counts based on the single passband. This is due to the fact that
galaxy interlopers and QSOs have been removed from the sample using
the $\chi^2$-technique applied to the objects detected in three or more
passbands. As illustration, the QSO number counts for the objects
originally classified as point sources is shown by the red histogram.

Beyond the morphological classification limit, the number counts based
on the single passband level drop to zero (by construction), while the
number counts based on the colour catalogue extend to significantly
fainter magnitudes. This illustrates the usefulness of using all
available colours to construct a point source catalogue. This is
especially true in $U$, where the counts reach almost two magnitudes
fainter than the single passband counts. This is because the $U$ data
is shallow compared to those in the other bands and most of the
information on the (P/E) classification comes from redder
passbands. The gain in magnitude is negligible in $R$ which, according
to Tab.~\ref{tab:infl}, in most cases defines the (P/E) classification.

\subsubsection{Comparison with the model and literature}

The counts predicted by the model and the observations from the colour
catalogue are in good agreement, considering that no fine tuning or
fitting was performed.  The observed data follow the thick black lines
within the errors indicating that the choice of \cs\ is appropriate
(or at least consistent with that predicted by the model). The data
fall below the dashed thick lines, which would be the predicted counts
if the (P/E) classification would be reliable for all
magnitudes. Based on the flattening of the observed counts and guided
by the results from the model the 90\% completeness limits are
estimated to be approximately $U$ = 23.8, $B$ = 24.0, $V$ = 23.5, $R$
= 23.0, $I$ = 21.0, $J$ = 20.5, $K$ = 19.0.

The deepest stellar counts available in the literature are from the {\sc
hst}, but they cover a small area, e.g. Santiago \etal (1996) combine
17 WFPC-2 fields for a total area of 0.027 deg$^2$, 90\% complete to
$I_{814}$ of 22 to 24 depending on the field.  At the other end of the
spectrum there is the recent analysis of Chen \etal (2001) using 279
deg$^2$ of {\sc sdss} data complete to a depth of $g^\prime$ = 21.0
(corresponding to approximately $V$ = 20.4 for a typical colour of
$g^\prime - r^\prime$ of 1, or three magnitudes brighter than ours),
or Ojha (2001) that analyse seven fields of {\sc 2mass} data covering
67 deg$^2$ down to $K_{\rm s}$ = 15.  Intermediate large area surveys
are e.g. the ``Deep Multicolor Survey'' (DMS; Osmer \etal 1998 and
references therein) covering six fields of about 0.14 deg$^2$ each
complete to about $B$ = 23.5 (comparable to ours), or Reid \& Majewski
(1993) covering 0.3 deg$^2$ complete to about $V$ = 21.5 (brighter by
two magnitudes compared to ours).

What will make the DPS survey unique when finished is that it will
cover a larger area, to a depth that is comparable or better, than
existing intermediate large area stellar surveys and that the data
will be in five or seven passbands.

\subsubsection{The slope of the IMF}
\label{cIMF}

\begin{figure*}[ht]
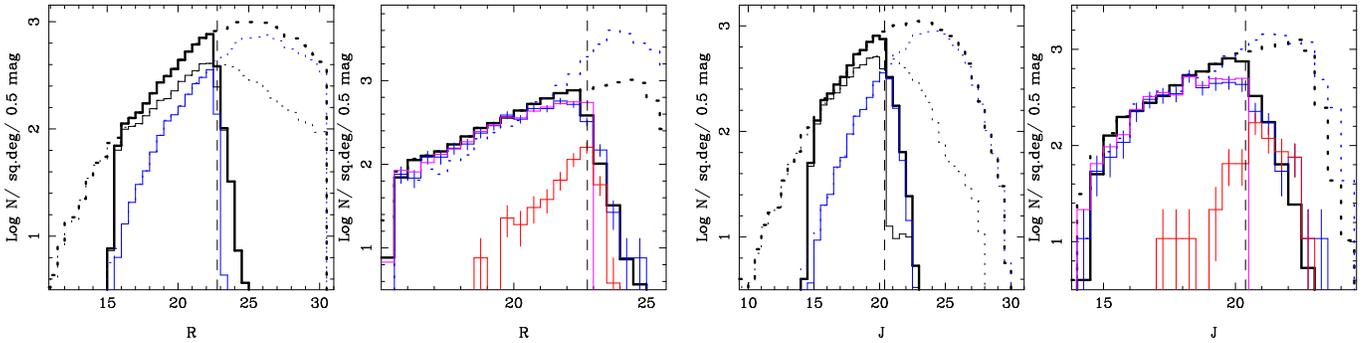

\begin{minipage}{0.49\textwidth}
\resizebox{\hsize}{!}{\includegraphics{MS2181f3a.ps}}
\end{minipage}
\hfill
\begin{minipage}{0.49\textwidth}
\resizebox{\hsize}{!}{\includegraphics{MS2181f3b.ps}}
\end{minipage}
\caption{
Number counts in $RJ$ with the power law form of the IMF from Chabrier (2001). 
The meaning of the curves is as in Fig.~\ref{fig:NC}. 
The counts at $R \sim 22$, $J \sim 19.5$ are about
20\% larger than for the standard case with the lognormal form.
}
\label{fig:NC_cIMF}
\end{figure*}

In this Section the influence of the shape of the IMF on the number
counts is illustrated.  Figure~\ref{fig:NC_cIMF} shows the number
counts in $R$ and $J$ when the power law form in Chabrier (2001) is
adopted instead of his lognormal law.  The number counts at $R \sim
22$ and $J \sim 19-20$ are larger than in the default case by about
20\%. In the other bands the effect is less striking but also present.
This illustrates the potential of the present data, especially when
the DPS survey is completed and data of similar depth on fields with
different galactic coordinates are available, to constrain e.g. the
IMF.  The counts in this field are better fitted with the lognormal
law proposed by Chabrier (2001) than the power law proposed by
him. The stars in the relevant magnitude interval have typical masses
of about 0.1 - 0.2 \msol. However, the fits are certainly not perfect,
and this possibly hints to a more complex, or different, shape of the
IMF than a power law or lognormal law.

\subsection{Colour-colour diagrams}
\label{CCD}

\subsubsection{General description}

\begin{figure}[ht]
\centerline{
\psfig{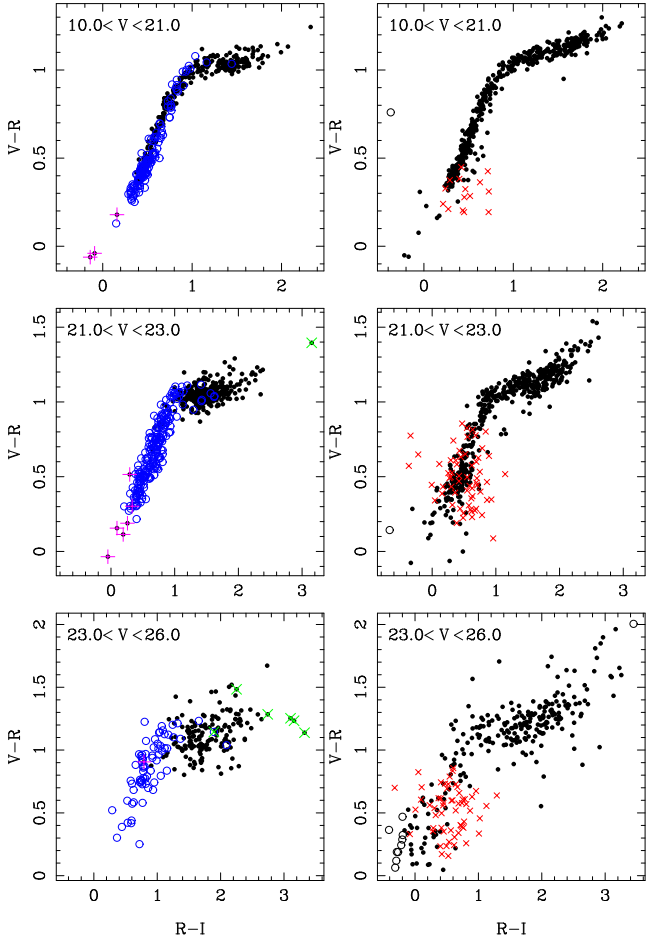}
}
\caption{
Colour-colour diagram for different cuts in magnitude. 
This $(V-R)-(R-I)$ diagram based on the five passband catalogue is a typical
example and the full figure is only available through the on-line edition.
Left panel: Model predictions. Halo (open circles) and disk (filled
circles), with stars with $\log g > 7$ (representing WDs) and
stars with masses $<$0.1 \msol\ (representing low mass stars)
additionally marked by a plus and cross sign, respectively. Since the
full simulation is run over $n-$times the observed area, only
$\frac{1}{n}$ of all model data points are plotted.
Right panel: the data. Open circles represent outliers, in the sense
that they are not within a certain magnitude difference of any of the
model data points in the left-hand panel (see text). Red crosses are
objects identified as QSOs by the $\chi^2$-technique.
}
\label{fig:CC_dia}
\end{figure}

Figure~\ref{fig:CC_dia} shows selected colour-colour diagrams split
into magnitude bins for the five and seven passband catalogues.  In
the left-hand panel the predictions by the model are shown.  Objects
from the halo (open circles) and the disk (filled circles), with stars
with $\log g > 7$ (representing WDs) and stars with masses
$<$0.1 \msol\ (representing low mass stars) additionally marked by a plus and
cross sign, respectively. The right-hand panel shows the stellar
objects (filled and open circles), while the objects identified as
QSOs by the $\chi^2$-technique are shown as crosses.

The overall agreement is very good in all cases and for all magnitude
bins. In particular the width of the sequences is well matched to the
data confirming a-posteriori the correctness of scaling the simulated
errors by SExtractor as was derived by the comparison to deeper
external data.

The largest discrepancies are: (a) in the $(V-R)-(R-I)$ diagram, where
for increasing $R-I$ the model predictions saturate at a level of
$V-R$ of $\sim 1$, while the observed data points continue to become
redder; (b) in the $(R-I)-(I-J)$ diagram, where the slope of the data
and the model differ for $I-J \more 0.8$; (c) in the $(V-J)-(J-K)$
diagram, where for $V-J >3$ the data seem to remain at a constant
colour, while there is a slope in the model predictions. Possible
reasons for the discrepancies are discussed in Sect.~\ref{FURTHER}.

\subsubsection{Outliers}
\label{outliers}

Obviously there are also outliers in the colour-colour diagrams on the
data side. These can be real (for example due to variability) or hint
to problems in the catalogue production. This issue is extensively
discussed in H2002. Most important for the present paper is to remind
that these outliers do not influence the assignment of the spectral
type as this is based on {\em all} available magnitudes and errors.

To guide the eye, outliers are marked by the open circles. Outliers
are defined in the present paper as follows\footnote{In H2002,
outliers were defined by considering only the data itself, using
dissimilarity measures. See H2002 for details.}. An imaginary box in
colour space is drawn around all of the model stars in the left-hand
panel, for each magnitude bin.  An object is defined an outlier if an
observed data point on the right is not within any of these boxes.
The width of the box was set after some experimentation to be 15-30\%
of the maximum of the range in both colours axis in a given magnitude
bin.

This concept is found useful to estimate the possible level of
remaining contamination by QSOs and unresolved galaxies that have not
been identified as such by the $\chi^2$-technique. Although the outliers
are marked in all colour-colour diagrams, the diagrams where QSOs are
most easily identified based on their difference in colours with
respect to normal stars are $(U-B)-(B-V)$ in five passbands and
$(U-B)-(B-V)$, $(B-V)-(V-K)$, $(V-J)-(J-K)$ in seven passbands
(H2002). Note, however, that the colour-colour diagrams indicate that
QSOs and stars overlap in colour-colour space, and therefore no
selection based purely on colour-colour diagrams can be made (see also H2002).

The number of outliers in the $(U-B)-(B-V)$ colour-colour diagram in
five passbands is 36, 28 of which are in the region occupied by QSOs,
while in the $(U-B)-(B-V)$, $(B-V)-(V-K)$, $(V-J)-(J-K)$ colour-colour
diagrams in seven passbands there are 63 outliers, 26 of
which are in the region occupied by QSOs.
However since stars do overlap with QSOs it can not be assumed
automatically that all outliers are QSOs.  The upper limits to the
remaining contamination by QSOs are therefore $<$6.3\% in the seven and
$<$2.3\% in the five passband catalogue. In Sect.~\ref{effectiveness} an
independent estimate of this contamination is given.

\subsection{Colour-magnitude diagrams}
\label{HRD}

\begin{figure}[ht]
\centerline{
\psfig{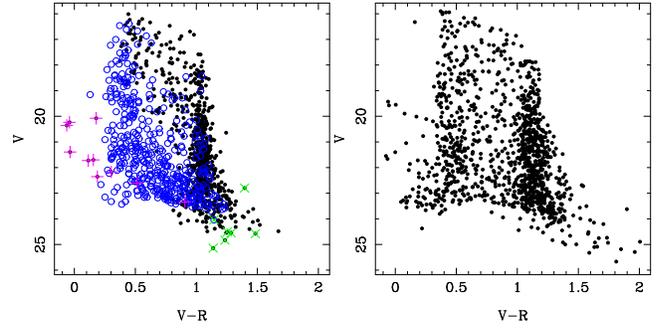}
}
\caption{
Colour-magnitude diagram. 
This ($V,V-R$) diagram based on the five passband catalogue is a typical
example and the full figure is only available through the on-line edition.
Left panel: model results. The symbols have the same meaning as in
Fig.~\ref{fig:CC_dia}. Since the full simulation is run over $n-$times
the observed area, only $\frac{1}{n}$ of all model data points are
plotted.  The right panel shows the data.
}
\label{fig:CM_dia}
\end{figure}

Figure~\ref{fig:CM_dia} shows selected colour-magnitude diagrams.  In
the left-hand panel the model predictions are shown with symbols as
introduced in Fig.~\ref{fig:CC_dia}. The right-hand panel shows the
data. The match is good, in particular the Galactic halo (with bluer
colours) and the Galactic disk (with redder colours) are readily
distinguished.

One can also notice the particular position occupied by the model very
low-mass stars (crosses), that are the reddest and faintest objects in
most plots. White dwarfs (plus signs), although found over a
relatively wide range of colours and magnitudes, are the only objects
to occupy a vertical strip to the blue of the bulk of halo stars: in
fact, in diagrams like $V$ vs.\ $V-R$ and $I$ vs.\ $V-I$, the sequence
of hot WDs (with $V-R<0.3$, $V-I<0.4$) starts at $V\ge20$ and
$I\ge20$, and this matches quite well the sequence of hot objects seen
in the observational plots.

\subsection{Colour distribution}
\label{CD}

\subsubsection{General description}

\begin{figure}[ht]
\centerline{
\psfig{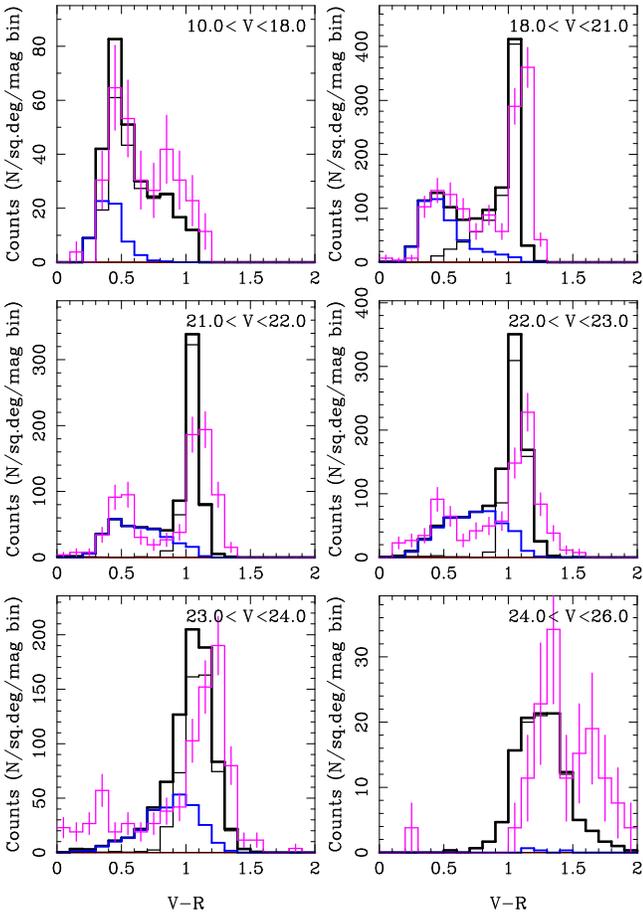}
}
\caption{
Colour distributions for different cuts in magnitude.  The full figure
is only available through the on-line edition.  Shown are the model
predictions (the halo is the blue lined histogram; the disk the thin
lined histogram, and their sum is the thick lined histogram) and the
data (the pink histogram with the error bars).}
\label{fig:Cd_dia}
\end{figure}

Although the colour-colour and colour-magnitude diagrams provide
useful and complementary information, the most stringent test for the
comparison between data and model is provided by the distributions
over colour at different magnitude slices, as provided in
Figure~\ref{fig:Cd_dia}.

The overall agreement is again very good.  The shape of the colour
distributions is correctly recovered in most cases; the absolute
number of objects in the faintest bin, however, is not always exact
(e.g. $U-B$ and $R-I$).

The most obvious discrepancies are: (a) the excess of objects with $-1
< U-B <-0.5$ and $22 < U < 24$ which could be QSOs not identified by
the $\chi^2$-technique, or WDs not predicted by the model; (b) the fact
that for disk stars the predicted $V-R$ colours are bluer than the
observed ones; (c) similarly, that for disk stars the predicted $R-I$
colours are bluer than the observed ones; (d) the excess of observed
sources with $J-K > 1.2$ and $18 < K < 19 $ which again could be QSOs, or
galaxies, not identified by the $\chi^2$-technique.

The shifts in colour seen in $V-R$ and $R-I$ are of the order 0.2 mag, but
the conclusion is that all the other colours are correctly predicted
at a level $\less 0.1$ mag.

\subsubsection{Further investigation of the model data}
\label{FURTHER}

\begin{figure}[ht]
\centerline{
\psfig{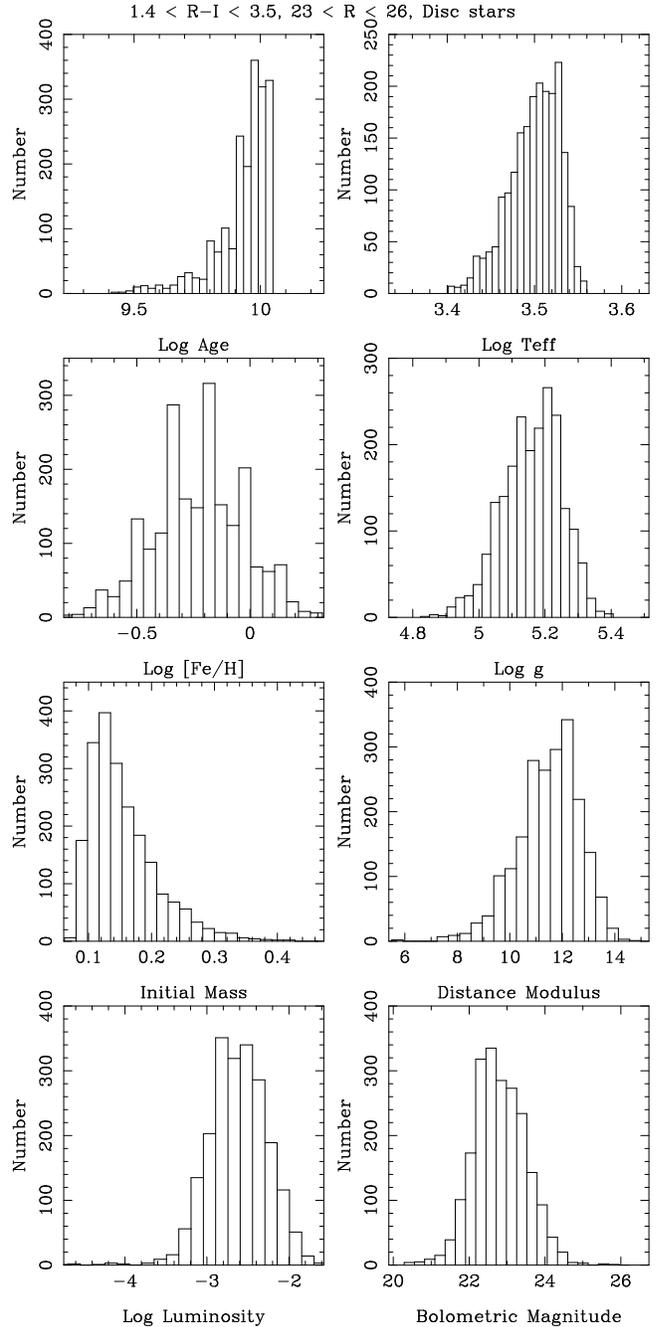}}
\caption{
Distributions over various parameters of all model stars which 
are formed in the disc and have $1.4 < R-I < 3.5$ and $23 < R < 26$.
}
\label{fig:VRRI}
\end{figure}

To make better use of the simulations a numerical code has been
developed that allows to search on any combinations of the parameters
provided by the Galactic Model convolved with the error model,
completeness, etc, as described earlier. The parameters provided are
the colours with their errors, the Galactic component (i.e. disk,
halo, bulge), age, [Fe/H], initial mass, luminosity, effective
temperature, $\log g$, distance modulus, visual extinction and
apparent bolometric magnitude.  Constraints on the differences between
two columns are allowed for as well. As an illustrative example the
problem of the discrepancy between model and data in the $(V-R)-(R-I)$
colour-colour diagram is addressed. Figure~\ref{fig:VRRI} shows the
distribution over various model parameters of all simulated stars that
are formed in the disc and fulfil $1.4 < R-I < 3.5$ and $23 < R < 26$.
From this analysis it is clear that mainly low mass stars ($\less$ 0.2
\msol) are contributing to this colour range indicating that likely
the theoretical colours for these stars are off by approximately 0.2
mag.

It should be pointed out that, among the possible inadequacies of the
model, systematic shifts in the model effective temperatures or in the
adopted stellar metallicities would affect all colours in more or less
the same way, which does not seem to be the case here.  In fact, just
a few of the colours are deviant, and this might be related to the
tables of bolometric corrections which are adopted in the model. For
instance, either these bolometric corrections were derived from
imperfect response curves for some of the EIS filters, or the
synthetic stellar spectra in use (see Girardi \etal 2002 for details)
present inadequacies in some of their wavelength regions (caused by
e.g.\ incomplete line opacity tables).  This matter will be further
investigated, but whatever is their cause, these colour problems are
small enough to not affect any of our conclusions.

Alternatively, these discrepancies may result from the data rather
than the model. It is important to point out that most of the
discrepancies occur at the near-infrared passbands ($IJK$) where the
data may be affected by fringing as well as other problems related
to the reduction of jittered infrared data.  It will be of interest to
see if these problems persist as the techniques for reducing WFI and
SOFI data are improved.

\subsection{The effectiveness of the $\chi^2$-technique }
\label{effectiveness}

In the selection of stellar sources, the values of the \cs\ of the
different passbands are combined as described previously. Then the
$\chi^2$-technique is used to filter out unresolved galaxies and QSOs. In
this section it is investigated how reliably the $\chi^2$-technique works,
and if it could be used as a stand-alone tool. 
 
The dashed blue histogram in Fig.~\ref{fig:NC} in fact shows the
number counts of objects that have been assigned ``stellar'' template
spectra and that are ranked 1 and $\le3$ when applying the
$\chi^2$-technique to the full, five, respectively seven, passband
colour catalogue, not just the point source catalogue\footnote{A
technical remark: The difference of selecting objects ranked 1 for
the five passband catalogue and ranked $\le$3 (i.e. all objects) for
the seven passband catalogue is related to the information provided by
the extra filters. More precisely, for objects that have an infrared
continuum described by a power law, such as low- to intermediate
redshift QSOs or hot WDs, the $J$ and $K$ magnitudes do not provide
any additional information about the SED but will impose additional
constraints, by increasing the number of the degrees-of-freedom in the
fitting. A possible answer to the problem of redundant information
could be a Principle Component Analysis. This will be investigated in
the future.}.

Even at relatively bright magnitudes these counts are significantly
larger than predicted by the model, in particular in $UBVR$. This is
very likely due to misclassifications, and cross-talk, between
galaxies and stars. Since the number density of galaxies is so much
larger than that for stars (already a factor of 10-15 at \msl), even a
10\% misclassification of star to galaxies and vice versa leads to a
very small decrease of galaxy number counts while it leads to a
doubling of the stellar number counts. This is illustrated further
below, and shows that the $\chi^2$-technique can not, and should not,
be used blindly.

\begin{figure}[ht]
%
\centerline{
\psfig{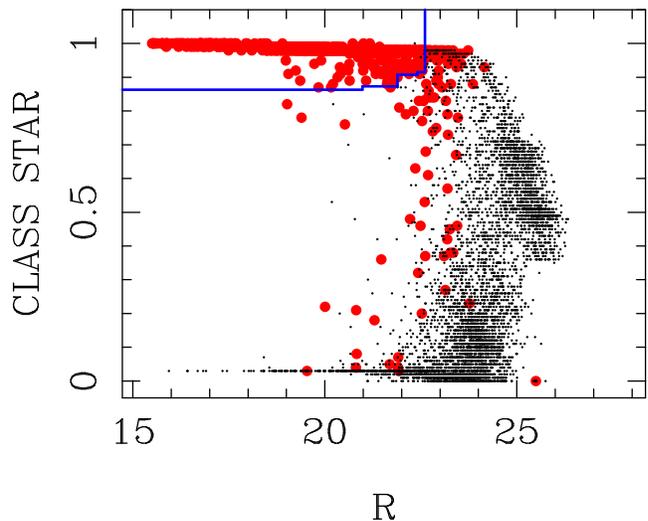}}
\caption{
\cs\ values of the objects that are classified as point 
sources in the five and seven passband colour catalogues (filled red
circles).  This panel shows the $R$-band based on the five passband
catalogue as an example, and the full figure is only available through
the on-line edition. Black dots represent the \cs\ values of the
objects that are classified as ``stellar'' and that are ranked 1 and
$\le3$ in the full five and full seven passband catalogue
respectively, when applying the $\chi^2$-technique. \msl\ and \cs\
limits are indicated.
}
\label{fig:cs_c}
\end{figure}

Fig.~\ref{fig:cs_c} shows the distribution of \cs\ in the different
bands for the objects that are classified point sources in the colour
catalogue (red circles). This gives an illustration of the usefulness
of using all available colours to classify objects. In $UV$ many
objects classified as point sources using all available colours are
outside the region that is initially set to contain point sources in
these bands individually. Table~\ref{tab:infl} contains information on
which band actually decided whether an object is classified as a point
source. This is $B,R$ and $I$ for the five passband catalogue, and
$B,R$ and $J$ for the seven passband catalogue. The dots in these
figures show the distribution of \cs\ for the objects assigned
``stellar'' template spectra that are ranked 1 and $\le3$ in the full
five and full seven passband catalogue respectively, when applying the
$\chi^2$-technique. It can be clearly seen that even bright objects
with small \cs\ values can be assigned a stellar spectral type, even
when these objects are verified to be resolved galaxies.

This illustrates the strength and weakness of using
\msl/\csl\ or the $\chi^2$-technique. The former method
selects point like objects without having the possibility to
distinguish between stars and QSOs or unresolved galaxies, while the
latter method uses astrophysical information (the SEDs of objects in
the Universe) to assign types but it does not consider whether an
object is resolved or not.

\begin{figure}[ht]
%
\centerline{
\psfig{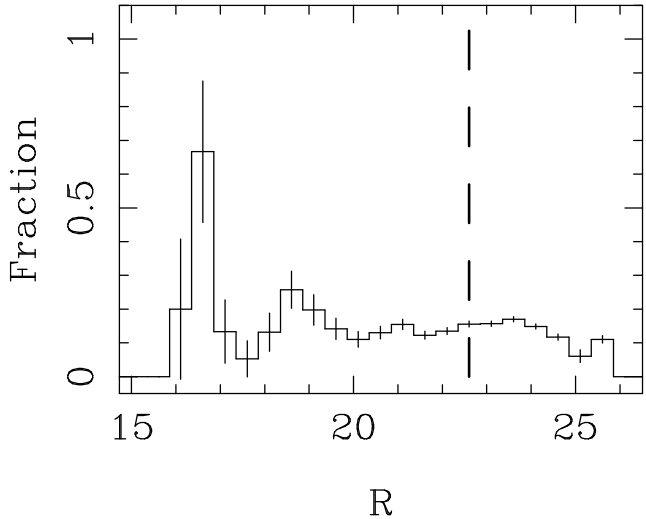}}
\caption{
Fraction of objects that are classified as ``stellar'' and that are
ranked 1 and $\le3$ in the full five and full seven passband catalogue
respectively, when applying the $\chi^2$-technique and that have a
\cs\ less than 0.1. \msl\ limits are indicated. This panel shows the
$R$-band based on the five passband catalogue as an example and the
full figure is only available through the on-line edition.
}
\label{fig:cs1_c}
\end{figure}

To investigate the level of misclassification of true galaxies as
stars by the $\chi^2$-technique, Fig.~\ref{fig:cs1_c} shows the ratio of
the number of objects assigned ``stellar'' spectra to all objects, for
values of \cs\ $<$ 0.1 (i.e. true galaxies). It is noticed that the
fraction of misclassification is essentially independent of magnitude
and ranges between 13 and 19\% in the five passband catalogue (with an
average of 16.3\%), and between 7 and 17\% for the seven passband
catalogue (with an average of 12.8\%). The smaller level of contamination 
indicates the effect of having more colours available in the fitting
to better constrain the assignment of the best fitting spectrum.

If it is assumed that the misclassification from true stars to
galaxies or QSOs is similar, one can estimate from the numbers in
Table~\ref{tab:infl} that 16.3\% out of the total of 173
QSO+unresolved galaxies, or 28 objects may be wrongly assigned such a
type, or that 28 stars are incorrectly removed from the stelar
catalogue. This constitutes 2.4\% of the sample of stars. The same
number is found for the seven passband catalogue.

\subsection{Spectral types and data tables}

Tables~\ref{tab:5PBTAB} and \ref{tab:7PBTAB} give the first entries of
the stellar catalogue in respectively five and seven bands. The tables
give the following information: in column (1) the EIS identification
number; in columns (2) and (3) the J2000 coordinates; in columns (4) -
(13) for the five passband and columns (4) - (17) for the seven passband the
magnitude (Vega system) and the SExtractor error (upper limits are
indicated by a $-1$; in column (14) (respectively column (18) in the seven
passband) the assigned spectral type, with the following meaning:
\begin{itemize}

\item Prefix ``MS''. Templates for normal O to M stars from the Pickles (1998) library. 
Stars that are detected in fewer than three bands and therefore have
not been fitted by the $\chi^2$-technique are listed as ``dummy'' here.

\item Prefix ``WD''. Templates for WDs from the models provided by D.~K\"oster 
(the numbers indicate effective temperature and $\log g$), Ibata \etal
(2000; the observed spectrum of F351-50, F821-07) or Oppenheimer
\etal (2001; the observed spectrum of WD 0346+246).

\item Prefix ``LMS''. Templates for Low Mass Stars from Chabrier \etal  (2000). 
The prefix is followed by a number indicating $T_{\rm eff}/100$ and
the type of model used (NeGen = Next Generation, AMESd = DUSTY, or
AMESc = CONDENSED, as defined in Chabrier \etal 2000).

\end{itemize}

\noindent
Tables~\ref{tab:5PBTABQSO}, \ref{tab:7PBTABQSO}, \ref{tab:5PBTABGAL}
and \ref{tab:7PBTABGAL} give the first entries of the likely QSOs and
unresolved galaxies in the five and seven passbands as identified by
the procedure in Sect.~\ref{Procedure}. The information listed is the same as
in the previous tables, except that in the last column the photometric
redshift is listed for the QSOs, and the type of galaxy (following
Coleman et al. 1980) and photometric redshift for the unresolved
galaxies as determined by the spectral template fitting method.

The complete tables can be retrieved from the CDS or from the URL
``http://www.eso.org/science/eis/eis\_pub/eis\_pub.html''.

\begin{table*}[ht]
\caption{First 15 entries of the stellar catalogue in five passbands.}
\psfig{figure=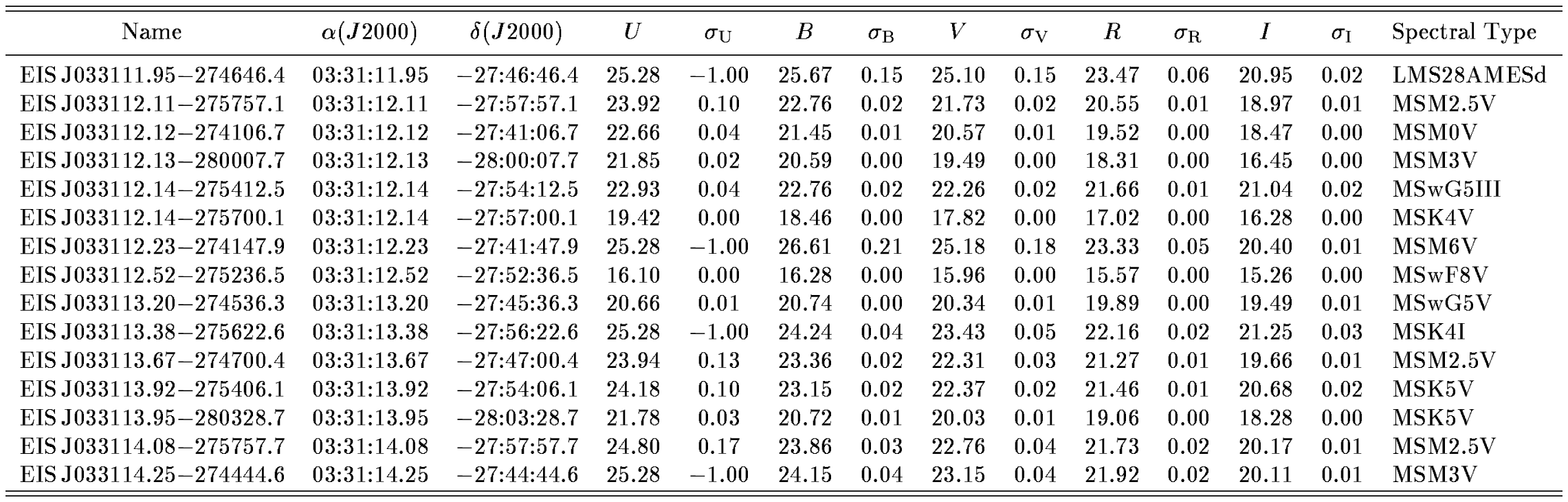,width=18.0cm,angle=-90}
\label{tab:5PBTAB}
\end{table*}

\begin{table*}[ht]
\caption{First 15 entries of the stellar catalogue in seven passbands.}
\psfig{figure=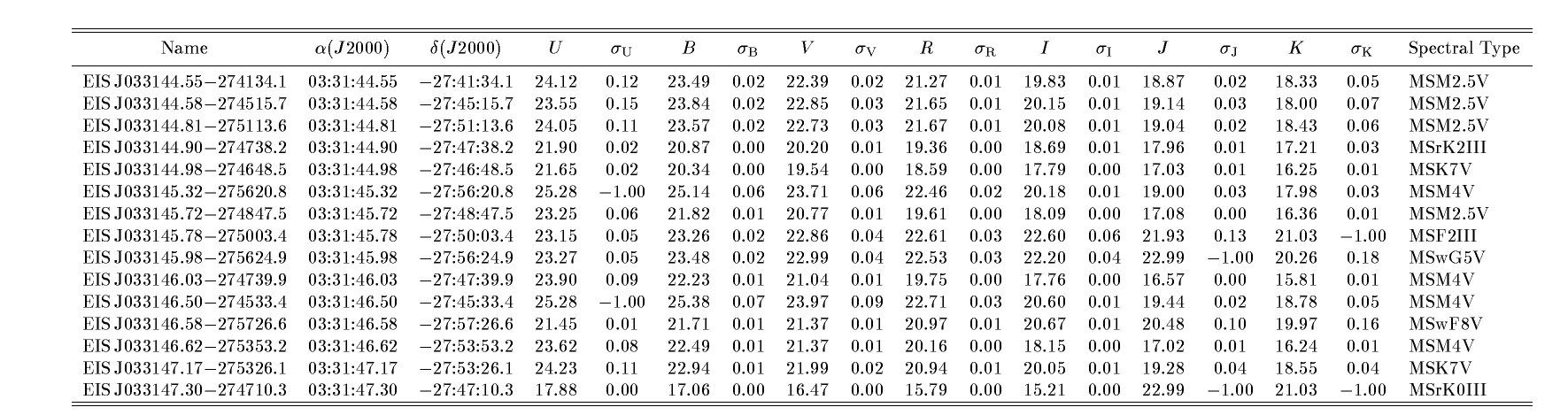,width=18.0cm,angle=-90} 
\label{tab:7PBTAB}
\end{table*}

\begin{table*}[ht]
\caption{First 15 entries of the likely QSOs in five passbands.}
\psfig{figure=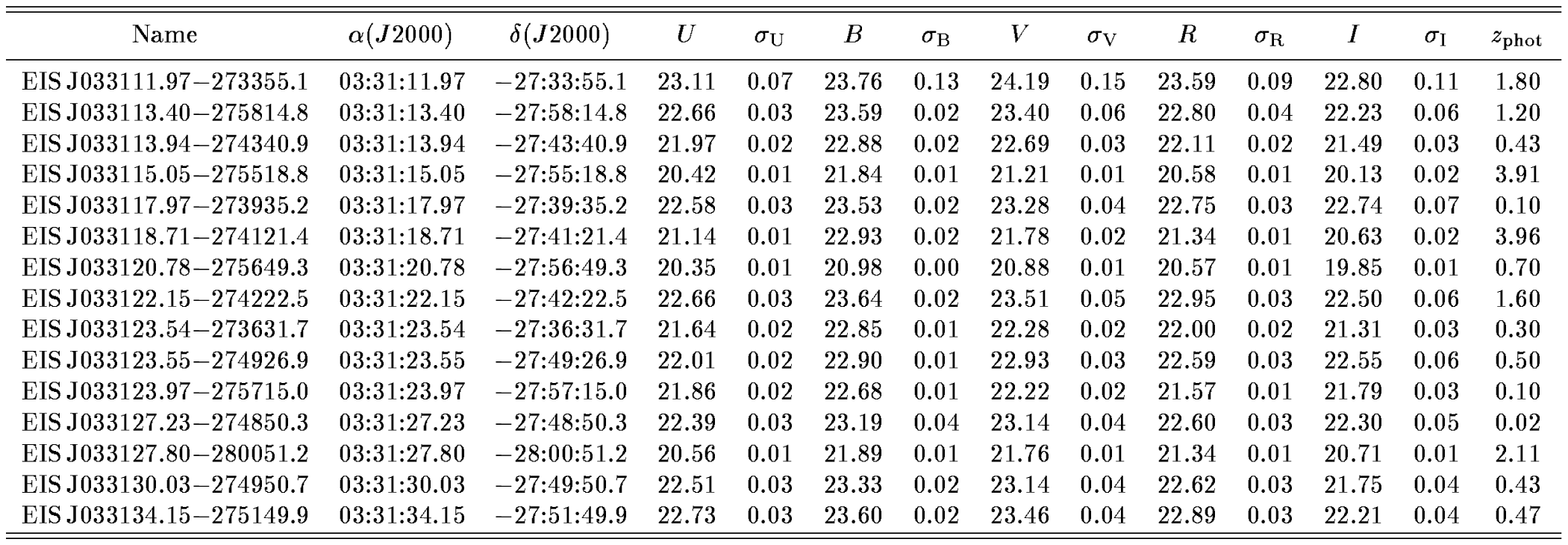,width=18.0cm,angle=-90} 
\label{tab:5PBTABQSO}
\end{table*}

\begin{table*}[ht]
\caption{First 15 entries of the likely QSOs in seven passbands.}
\psfig{figure=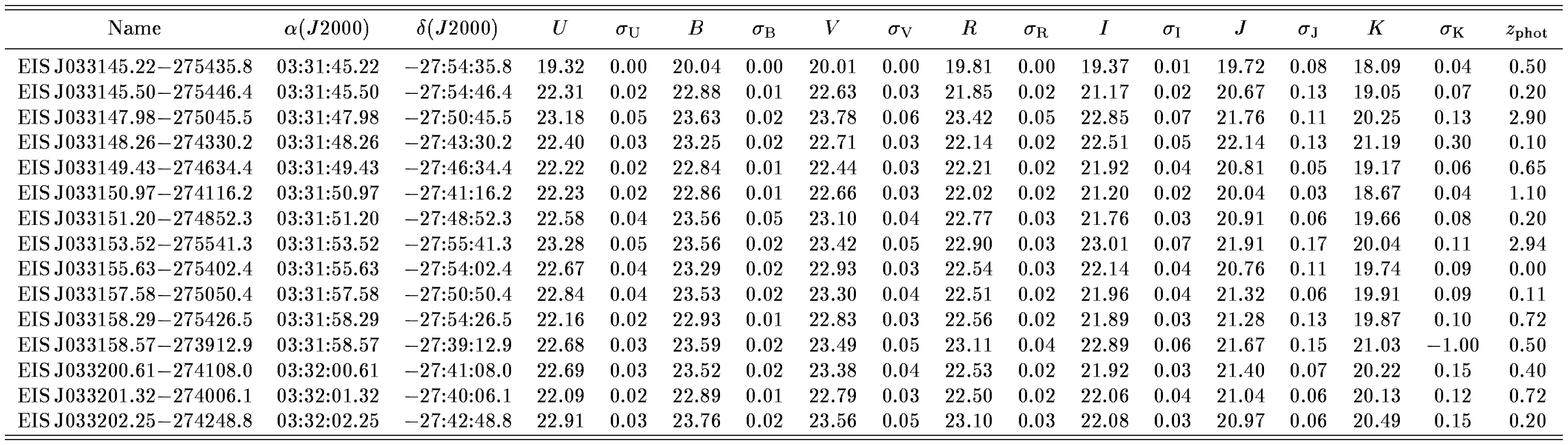,width=18.0cm,angle=-90} 
\label{tab:7PBTABQSO}
\end{table*}

\begin{table*}[ht]
\caption{First 15 entries of the likely unresolved galaxies in five passbands.}
\psfig{figure=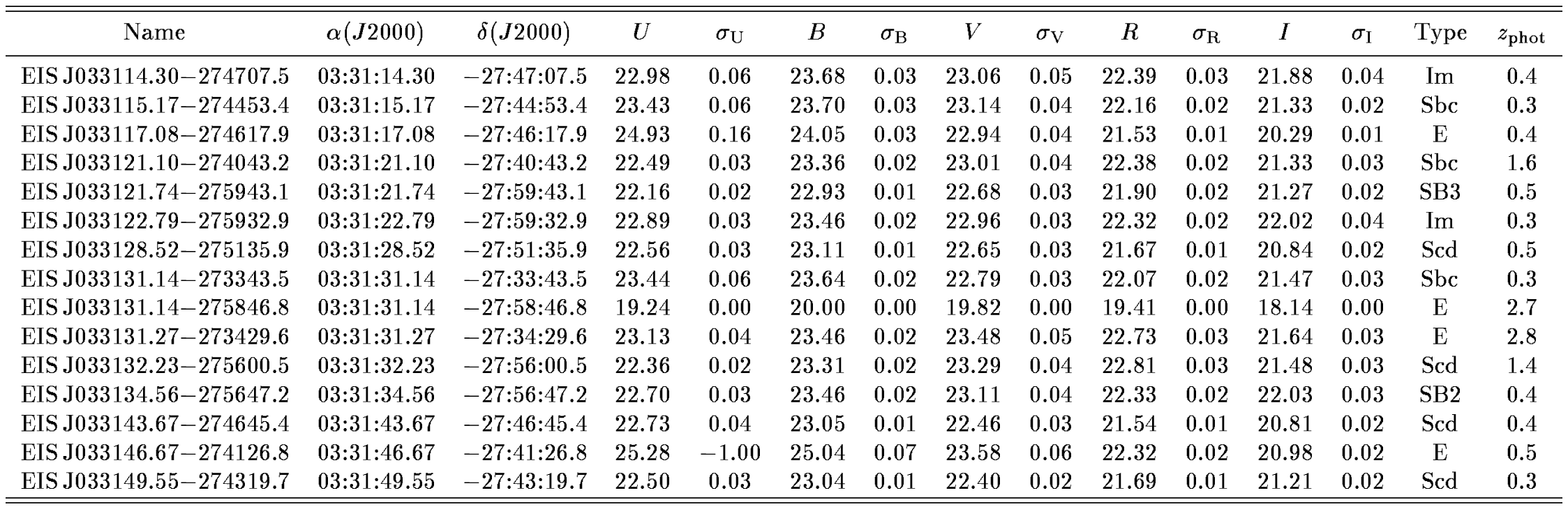,width=18.0cm,angle=-90} 
\label{tab:5PBTABGAL}
\end{table*}

\begin{table*}[ht]
\caption{First 15 entries of the likely unresolved galaxies in seven passbands.}
\psfig{figure=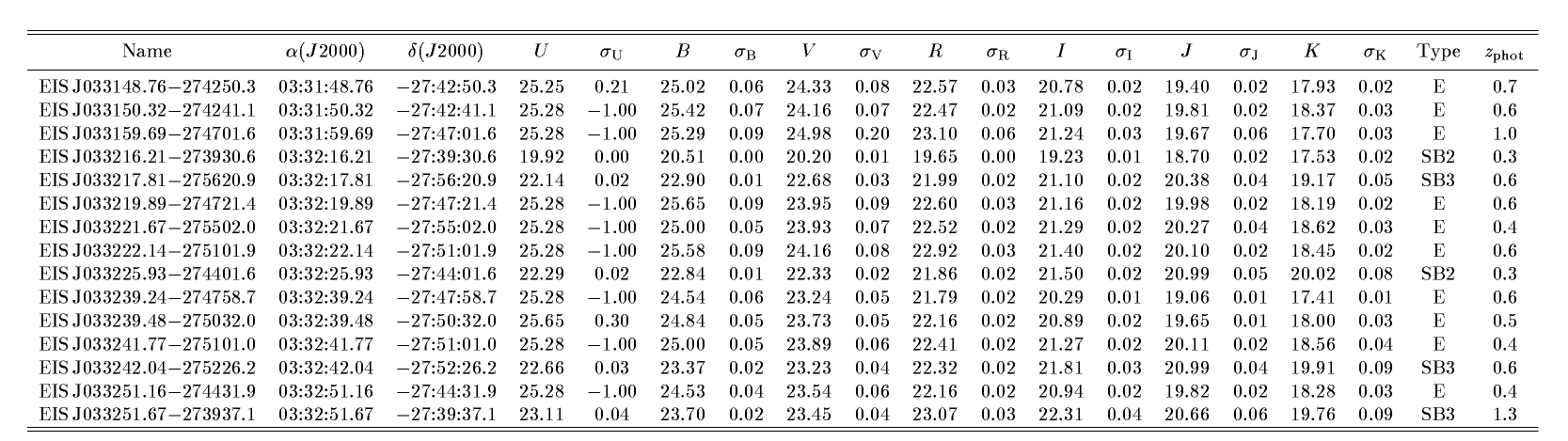,width=18.0cm,angle=-90} 
\label{tab:7PBTABGAL}
\end{table*}

\section{Summary and conclusions}
\label{summary}

This paper describes the procedures adopted in the construction of
multi-colour stellar catalogues, suitable for statistical studies,
extracted from multi-band imaging data. The procedure involves several
steps among which: a magnitude-dependent scheme to morphologically
classify sources identified using SExtractor; a procedure to combine
the information available in catalogues extracted from images taken in
different passbands; and the use of a $\chi^2$-technique to
assign spectral types to the sources using the multi-band
information. This allows to minimise the contamination of the stellar
catalogue by QSOs and unresolved galaxies, and to assign spectral types
for the stars, which are included in the stellar catalogues presented
in this paper.

The methodology outlined here is applied to the first data set
released for the DPS, the CDF-S. The 90\% completeness limits are
estimated to be approximately $U$ = 23.8, $B$ = 24.0, $V$ = 23.5, $R$
= 23.0, $I$ = 21.0, $J$ = 20.5, $K$ = 19.0.  To assess the quality of
the catalogues produced, number-counts, colour-colour and
colour-magnitude diagrams, and colour-distributions derived from the
data are compared to those obtained from simulated catalogues. 
Mock catalogues are created using a Galactic model based on population
synthesis (described by parameters set by independent data), an
error model describing the expected photometric errors, saturation and
completeness as derived from the data. Even though no attempt is made
to fine tune the Galactic model parameters the agreement between real
and simulated data is remarkable, serving as a good indicator of the
reliability of the catalogues produced. The comparison also suggests
that the depth of the data is suitable to constrain the IMF and/or SFR
of low-mass stars.

This paper represents a first attempt to define procedures to produce
well-defined, deep stellar catalogues with minimal contamination by
QSOs and unresolved galaxies. The results are encouraging and
demonstrate the valuable contribution that the homogeneous multi-band
optical/infrared data set from DPS, probing different directions of
the Galaxy, will make when completed.

\begin{acknowledgements}

  L.G. thanks ESO for the kind hospitality during two visits.

\end{acknowledgements}

\appendix

\section{Determining \msl\ and \csl}

\begin{figure}[htb]
\centerline{ \psfig{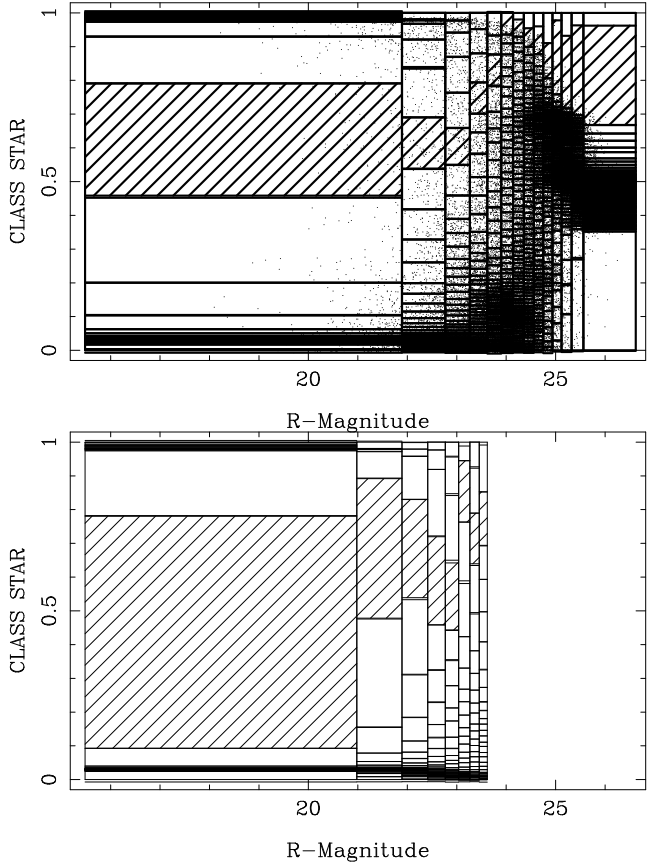} }
\caption{\cs\ as a function of $R$-band magnitude, with the binning in magnitude 
and \cs\ indicated, for the first pass (top), and second pass (bottom,
without data points for clarity). The hatched bin is the one that
defines the background of galaxies.  }
\label{fig:binning}
\end{figure}

This appendix gives a brief technical description of the way \csl\ and
\msl\ are determined in the present paper. Refinements of this scheme 
might be considered in the future after completion of the DPS survey
(12 times the area under consideration here in different directions)
and analysis of their respective stellar catalogues, spectroscopic
follow-up studies that indicate the (in)correctness of the assigned
spectral types, or more extensive numerical simulations including
galaxies and QSOs.

One way to analyse a diagram like Fig.~\ref{fig:CS} is to set two
levels of significance. The first, $t_1$, setting, for a fixed
magnitude range, the level of significance of detecting stars against
the background of galaxies at some value of \cs, and $t_2$, setting
the level of significance of detecting stars at some magnitude, given
a fixed range in \cs.

This idea is implemented in the following way. The first step is to
allow for the single passband catalogue to be cut at a magnitude error
level, and keeping only objects with ``good'' SExtractor flags.

For the present analysis, only objects with SExtractor flag $<$4 are
kept and the default cut of S/N $>2$ is imposed.  The objects retained
are divided in bins of magnitude and \cs.  Instead of using a fixed
bin size in magnitude and \cs\ the bins are constructed in such a way
that they contain an equal number of objects, that is, the width of
the bin becomes the variable, rather than the number of objects in a
bin of fixed size.  The number of bins used is set automatically and
is based on the total number of  objects.

This is illustrated in Fig.~\ref{fig:binning} (top panel) for the
$R$-band, where the objects are divided into 14 bins in magnitude and
53 bins in \cs\ in the first pass. When the density of objects is high
the individual bins in \cs\ are no longer discernible.

For each magnitude bin, the distribution over \cs\ is analysed in the
following way. The bin with the largest width in \cs\ is considered to
define the background of galaxies (indicated by the hatched area in
Fig.~\ref{fig:binning}). A correction is made for the number of
galaxies relative to the total number of objects in that bin based on
the ratio of the number of objects below this bin to the total number
of objects. For simplicity it is assumed that the background level is
constant for the bins at higher \cs.  The mean and rms (based on
$\sqrt{N}$ statistics) per unit \cs\ are determined. This is a
necessary step as all bins have a different width.  For the bins that
have a \cs\ value higher than that defining the background bin, the
background level expected for that bin width is subtracted and divided
by the rms level expected for that bin width. This gives a
significance level per bin, ${\sigma}_{\rm bin}$.  For an external
threshold value, $t_1$, the lowest bin that has a significance above
this threshold is determined and, by linear interpolation using the
next highest bin, the final value for \csl\ for that magnitude bin is
calculated. In addition, all the bins that have a \cs\ larger than the
bin that defines ${\sigma}_{\rm bin}$ are combined and a similar
significance level for the group of stars as a whole (${\sigma}_{\rm
group}$) is computed.
Once the loop over the magnitude bins is completed, and given a second
external threshold level, $t_2$, that magnitude bin is determined
where $t_2 > {\sigma}_{\rm group}$. By linear interpolation using the
next brightest magnitude bin, the final value for \msl\ is computed.
Additionally, the magnitude, $m_{\rm cut}$, is determined below which
stars can not be recognised with any confidence. Then, for a bin in
magnitude to the left of \msl\ (since \msl\ signifies a right-sided
cut-off) and containing an equal number of objects as before, the
binning in \cs\ and the analysis to derive the value for \csl\ is
repeated.  This whole procedure is done twice, the first time using
all objects that have passed the selection on magnitude error and
SExtractor flag, and a second pass only retaining objects brighter
than $m_{\rm cut}$.  This allows for a better sampling over the
magnitude interval where stars can be identified.  The binning for the
second pass and the effect of only retaining objects brighter than
$m_{\rm cut}$ is illustrated in the bottom panel of Fig.~\ref{fig:binning}.
Figure~\ref{fig:CS} shows for the $R$-band the distribution over \cs\
and \csl\ and \msl\ for the adopted choice of $t_2 = 30$ and $t_1 =80$.

An independent check on the scheme presented, is to consider galaxy
counts as the complement to the objects selected to be point sources,
the argument being that if one is too liberal in the choice of \msl\
and \csl\ to classify stars, the remaining number of ``non-stars''
could be inconsistent with known galaxy counts. In $B$, $R$ and $K$
the number of observed objects in the colour catalogue not classified
as stars in the 0.5 mag bin below the faintest derived \msl\ limit
(i.e. $23.35 < B < 23.85$, $22.25 < R < 22.75$, $18.30 < K < 22.80$)
are, respectively, 2209, 1935 and 389.  The number density of galaxies
quoted in the literature is between 6500-11000, 5000-9000 and
4000-8000 per deg$^2$ per 0.5 magnitude bin in $BRK$ respectively (see
the extensive compilation at
http://star-www.dur.ac.uk/$\sim$nm/pubhtml/counts/counts.html for
detailed references). The predicted number of galaxies for the
effective area is 1710-2893, 1315-2367 and 371-742 in $BRK$
respectively, in agreement with the numbers observed, and indicating
that the choice of \msl\ and \csl\ to classify stars is not too liberal.

\end{document}